%% file: elsarticle-template-num.tex
\documentclass[review,3p,times]{elsarticle}

\usepackage{setspace}
\usepackage{amssymb}
\usepackage{amsmath}
\usepackage{bm}
\usepackage{siunitx}
\usepackage{setspace}
\usepackage{float}
\usepackage{graphicx}
\usepackage{placeins}
\usepackage{lineno}
\usepackage{cleveref}
\usepackage[numbers]{natbib}

\makeatletter
\def\ps@pprintTitle{%
   \let\@oddhead\@empty
   \let\@evenhead\@empty
   \let\@oddfoot\@empty
   \let\@evenfoot\@empty}
\makeatother

\usepackage{soul}       
\usepackage{xcolor}     

\journal{Powder Technology}
\input{commands}

\begin{document}

\newcommand{\overlap}{\delta_{\mathrm{n}}}
\newcommand{\tdisplacement}{\vm{\delta}_{\mathrm{t}}}
\newcommand{\me}{m_{\mathrm{e}}}
\newcommand{\muT}{\mu_{\mathrm{t}}}
\newcommand{\muTE}{\mu_{\mathrm{t,e}}}
\newcommand{\muR}{\mu_{\mathrm{r}}}
\newcommand{\muRE}{\mu_{\mathrm{r,e}}}
\newcommand{\Ee}{e_{\mathrm{e}}}
\newcommand{\Ree}{R_\mathrm{e}}
\newcommand{\CRD}[1]{$\mathrm{CRD}_{#1}$}
\newcommand{\LRD}[1]{$\mathrm{LRD}_{#1}$}
\begin{frontmatter}

\title{From Angle of Repose to Heap Morphology: Full-Field Calibration of DEM for Granular Powders}

\author[1,2]{Olivier Gaboriault}
\author[1]{Antonella Succar}
\author[1]{Cléo Delêtre}
\author[3]{Anatolie Timercan}
\author[3]{Roger Pelletier}
\author[2]{David Melancon} 
\author[1]{Bruno Blais\footnote{Corresponding author.
Email address: bruno.blais@polymtl.ca (Bruno Blais)}}

\affiliation[1]{organization={Department of Chemical Engineering, High-performance Automatization Optimization and Simulation (CHAOS) laboratory, Polytechnique Montréal},
            addressline={2500 Chemin de Polytechnique}, 
            city={Montréal},
            postcode={H3T 1J4}, 
            state={Québec},
            country={Canada}}
\affiliation[2]{organization={Department of Mechanical Engineering, Laboratory for Multiscale Mechanics (LM2), Polytechnique Montréal},
            addressline={2500 Chemin de Polytechnique}, 
            city={Montréal},
            postcode={H3T 1J4}, 
            state={Québec},
            country={Canada}}
\affiliation[3]{organization={National Research Council Canada},
            addressline={75 Bd de Mortagne}, 
            city={Boucherville},
            postcode={J4B 6Y4}, 
            state={Québec},
            country={Canada}}




\begin{abstract}
The calibration of discrete element method (DEM) models is commonly performed by tuning model parameters to match an experimental measurements, most commonly the angle of repose (AOR). Although widely used, AOR-based calibration metrics do not adequately characterize the full heap morphology, particularly when dealing with cohesive granular materials. As a result, AOR-based calibrations often leads to non-unique parameter sets. In this work, we propose a DEM calibration procedure based on full-field image analysis of static powder heaps rather than scalar AOR measurements. The method compares an average experimental heap profile (AEHP), obtained from repeated GranuHeap\texttrademark{} experiments, with an average numerical heap profile (ANHP) generated from DEM simulations. This comparison is performed using pixel-wise grayscale intensity values of both average heap profiles. Two metal powders commonly used in additive manufacturing, Ti6Al4V and Al6061, are used to evaluate the proposed methodology. This work highlights the limitations of traditional AOR-based approaches and demonstrates that full-field heap morphology offers a more reliable framework for DEM calibration.
\end{abstract}

\end{frontmatter} 

\section{Introduction}
\label{sec:introduction}
Granular matter, like powders, are used for a wide range of applications. For example, in additive manufacturing, metal and polymer powders can be used to fabricate functional parts that are difficult or impossible to build using conventional manufacturing techniques \cite{sames2016}. In the pharmaceutical industry, powders are frequently used as part of drugs encapsulation and drug formulation \cite{jadidi2025}. Finally, the food industry relies heavily on the storage and processing of granular ingredients like flour and grains. Given the widespread use of powders in industrial processes, predicting the performance and dynamics of granular material during their handling and manipulation is critical. In recent years, the discrete element method (DEM) has proven to be an efficient tool to achieve such prediction.

The DEM is a numerical method introduced by Cundall and Strack \cite{cundall_1979} that simulates granular systems such as the powder spreading step for additive manufacturing \cite{yim2022, yim2025a, wu2026, gaboriault2026}, granular mixers \cite{herman2022}, fluidized beds \cite{ferreira2023} and many more. This method consists in tracking the position and velocity of particles over time by computing forces and torques resulting from particle-particle and particle-wall interactions derive from different models (e.g., contact forces, sliding friction, rolling friction, cohesive forces, etc.), requiring their own parameters (e.g., damping coefficient, sliding friction coefficient, rolling friction coefficient, surface energy, etc.). Using these forces and torques, Newton's second law of motion and Euler's law of angular motion are solved numerically for each particle using an explicit time integration scheme. As a result, the DEM provides a high level of detail, enabling granular systems to be studied at the particle scale, but requires substantial computational resources compared to other methods \cite{chen2016, tian2018, golshan2023}. 

When simulating granular systems, the choice of interaction models and the values of their parameters have a drastic effect on their overall behaviour. The model parameters selection is usually achieved by comparing simulations and experiments of small-scale test cases for the same particles. Model parameters are then tuned, manually or automatically, to match a certain metric like the torque applied by the particle on a shaft \cite{werner2025} or the angle of repose (AOR) of a static \cite{wu2026, meier2019a, forgber2022} or dynamic \cite{yim2022} particle heap. This calibration process has many drawbacks as it requires a large number of tuning simulations---thus high computational resources---and does not guarantee a unique set of model parameters.  Furthermore, wall-related model parameters are often assigned constant values or simply assumed to be equal to the particle–particle parameters, which represents a strong assumption and can introduce significant uncertainty in the calibration process. These limitations become even more critical for cohesive powders, where complex interparticle interactions make scalar metrics such as the AOR insufficient to fully characterize the bulk behaviour and often lead to non-unique calibration solutions.

In this work, we propose a novel procedure based on the image analysis of static powder heaps shapes. Rather than relying on a scalar metric such as the AOR, the entire heap profile is used to evaluate the adequacy of a given combination of model parameters in comparison with experimental observations. First, we introduce the calibration procedure and assess the reproducibility of the heap experiments. Then, we present the procedure to calibrate two metal powders used in additive manufacturing. The results of the calibration are compared with those obtained using more traditional metrics, which rely on the AOR. We show that different clusters of calibration parameters emerge depending on the calibration procedure and the definition of the AOR used. We conclude by discussing future work which aims at disambiguating the parameters set in DEM simulations for cohesive powders.

\section{Methodology}
\label{sec::experimental}
The DEM calibration procedure that we introduce in this work relies on the comparison of static granular heap profiles that are obtained experimentally and from DEM simulations. This comparison is performed using the pixel values of two grayscale images: one representing the Average Experimental Heap Profile (AEHP) obtained from the granular material of interest and the other representing a numerically simulated heap profile generated using a given set of DEM model parameters. The current section first presents the powders used to demonstrate the proposed calibration procedure, followed by the experimental methodology employed to generate the static heap, which is schematized into a flow chart. Finally, the section conclude on the numerical framework used in our DEM simulations.

\subsection{Powders}
\label{subsec:powders}
\begin{figure}[!ht]
        \centering
        \includegraphics[width=90mm]{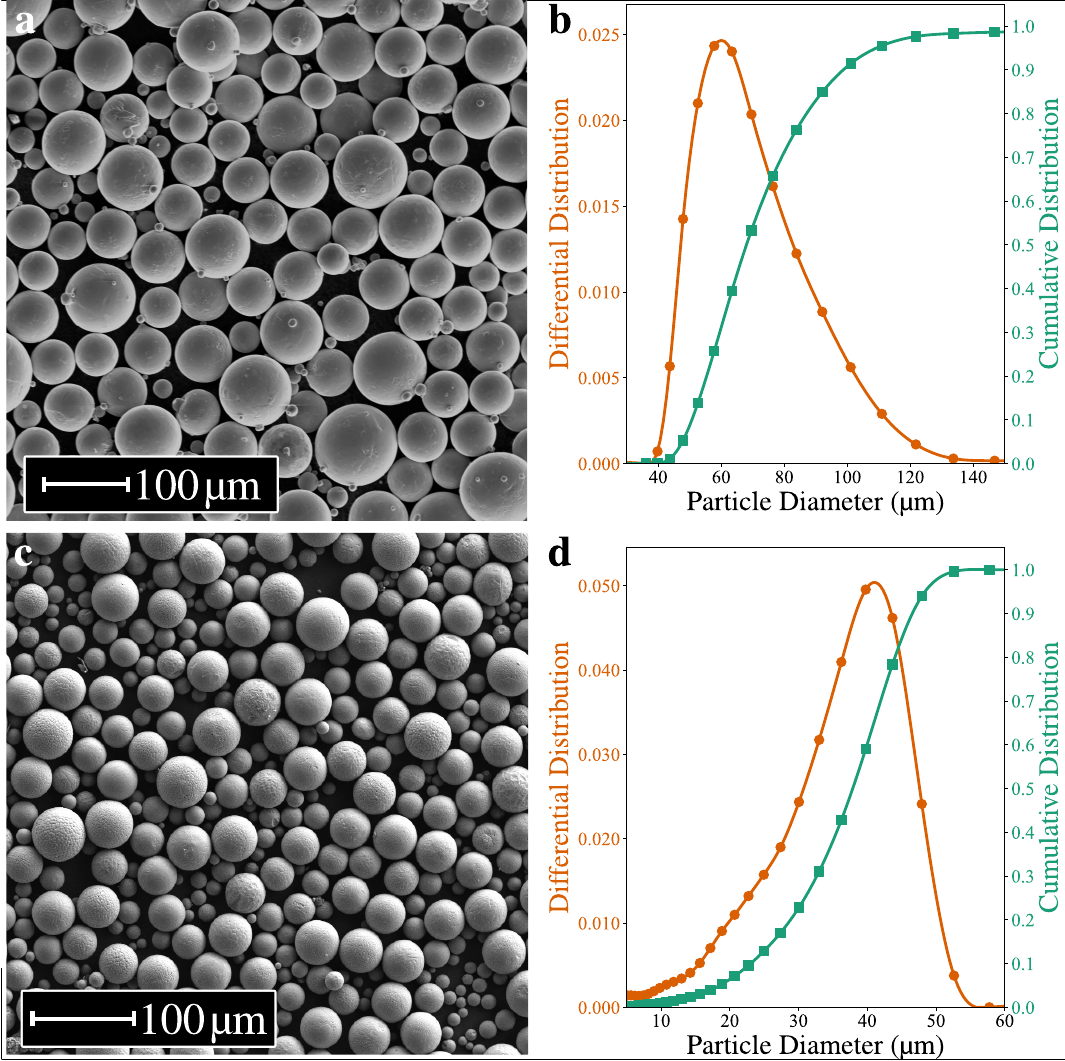}
        \caption{\textbf{Morphology and particle size distributions of Ti6Al4V and Al6061 powders.} \textbf{(a)} Scanning electron microscopy (SEM) image of the Ti6Al4V powder from Advanced Powders and Coatings (AP\&C) showing predominantly spherical particles with few satellites taken with a NANOS tabletop SEM from Semplor at 15 kV. \textbf{(b)} Corresponding differential (orange) and cumulative distribution (green) of the Ti6Al4V powder based on volume. \textbf{(c)} SEM image of the Al6061 powder from Equispheres showing spherical particles with some surface texture. \textbf{(d)} Corresponding differential (orange) and cumulative distribution (green) of the Al6061 powder based on volume.}
        \label{fig::powders}
\end{figure}

Two powders are used in this work to demonstrate the efficacy of our calibration method. The first one is a Ti6Al4V powder from Advanced Powders and Coatings (AP\&C) while the second one is a Al6061 powder from Equispheres. Figure~\ref{fig::powders} presents scanning electron microscopy (SEM) images and volume-based particle size distributions (PSD) for each powder. The titanium powder present predominantly spherical particles with few satellites (Figure~\ref{fig::powders}a), with a PSD ranging from 45 to 111~\si{\micro\meter} for the 5th–95th percentile ($D_{5}$-$D_{95}$) (Figure~\ref{fig::powders}b). The aluminium powder consists of spherical particles with some surface texture (Figure~\ref{fig::powders}c) with a PSD ranging from 18 to 50~\si{\micro\meter} for the $D_{5}$-$D_{95}$ (Figure~\ref{fig::powders}d). 

\subsection{Experimental methodology}
The flow chart for the experimental methodology is illustrated in Figure~\ref{fig::Experimental_flow_chart}. While any static heap experiment could be used with the proposed calibration procedure, we employ a commercial device, the GranuHeap\texttrademark \cite{yablokova2015}, which characterizes powder flowability through a controlled heap formation process. The device consists of a cylindrical container filled with powder that is vertically lifted by a robotic arm at a prescribed velocity, allowing the granular material to flow downward under gravity and form a conical heap on a base plate. As the cylinder rises, the powder flows and reorganizes until a stable heap is obtained. For each calibration run, the heap is photographed 16 times (Figure~\ref{fig::Experimental_flow_chart}a) over a $\SI{180}{\degree}$ arc around the static heap. The images are binarized to clearly distinguish the heap from the background. The relevant experimental dimensions of the apparatus, as well as the prescribed lifting velocity used in this study, are presented in Table~\ref{tab::experimental_parameter}.

\begin{table*}[!hbp]
    \caption{\textbf{GranuHeap experimental parameters.}}
    \centering
    \begin{tabular}{c | c}
        Parameter & Value \\
        \hline
        Cylinder inner diameter ($D_i$) & $\SI[minimum-decimal-digits=2]{1.0}{\centi\meter}$ \\
        Base plate diameter ($D_b$)     & $\SI[minimum-decimal-digits=2]{1.0}{\centi\meter}$ \\
        Lifting velocity ($v_l$)        & $\SI[minimum-decimal-digits=2]{1.0}{\centi\meter\per\second}$ \\
    \end{tabular}
    \label{tab::experimental_parameter}
\end{table*}

\begin{figure}[!htbp]
        \centering
        \includegraphics[width=75mm]{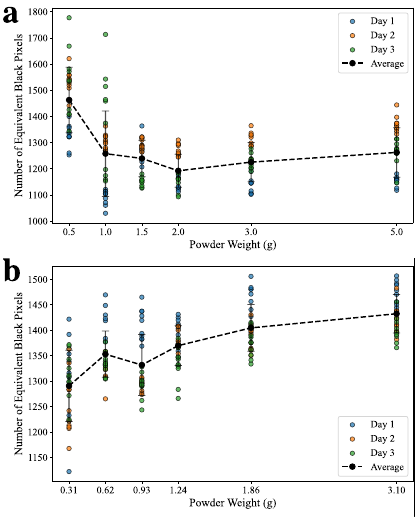}
        \caption{\textbf{Powder weight effect on number of equivalent black pixel.} Number of Equivalent Black Pixels (NEBP) as a function of the powder weight used during the GranuHeap experiment for 
        \textbf{(a)} titanium and \textbf{(b)} aluminium powders. The day on which the experiment was performed is identified by blue (Day 1), orange (Day 2), and green (Day 3) markers. The average and standard deviation over the 30 experiments are indicated by black markers with corresponding error bars.}
        \label{fig::NEBP_Weight}
\end{figure}

The heap profile obtained with the GranuHeap is sensitive to the amount of powder used inside the cylindrical container, up to a certain quantity. For instance, using an insufficient amount of powder results in a heap that is not fully formed and that is highly sensitive to particle placement in the container before lifting, leading to significant variability in the resulting heap profiles and poor repeatability between calibration experiments which is undesirable. On the other hand, we want to limit as much as possible the amount of powder used inside our cylindrical container to ensure that the corresponding simulations are not too computationally intensive. Indeed, an excessive powder volume would result in a larger number of particles, thus in a higher computational cost per simulation. With this taken into consideration, the first step of our calibration methodology is to determine the adequate amount of raw material to use for a given powder. To do so, we repeat the GranuHeap experiment 30 times (Figure~\ref{fig::Experimental_flow_chart}a-b) for predetermined powder weights, visible in Figure~\ref{fig::NEBP_Weight} for each powder type. Those experiments are performed on three different days, with 10 repetitions per day for a given powder mass, to account for environmental variability such as changes in air humidity that can affect the powder rheology \cite{ashrafizadeh2026,cordova2020}. For each run, an average heap profile is generated by averaging its 16 images. In other words, the 16 binary images of a run (Figure~\ref{fig::Experimental_flow_chart}a), which have pixel values that are 0 or 1, are converted into one grayscale image (Figure~\ref{fig::Experimental_flow_chart}b), which have pixel intensity values ($I_p$) between 0 (black) and 255 (white) inclusively. Note that black signifies the presence of powder. From these grayscale images, we compute the total number of equivalent black pixels ($NEBP$): 

\begin{figure}[!ht]
        \centering
        \includegraphics[width=160mm]{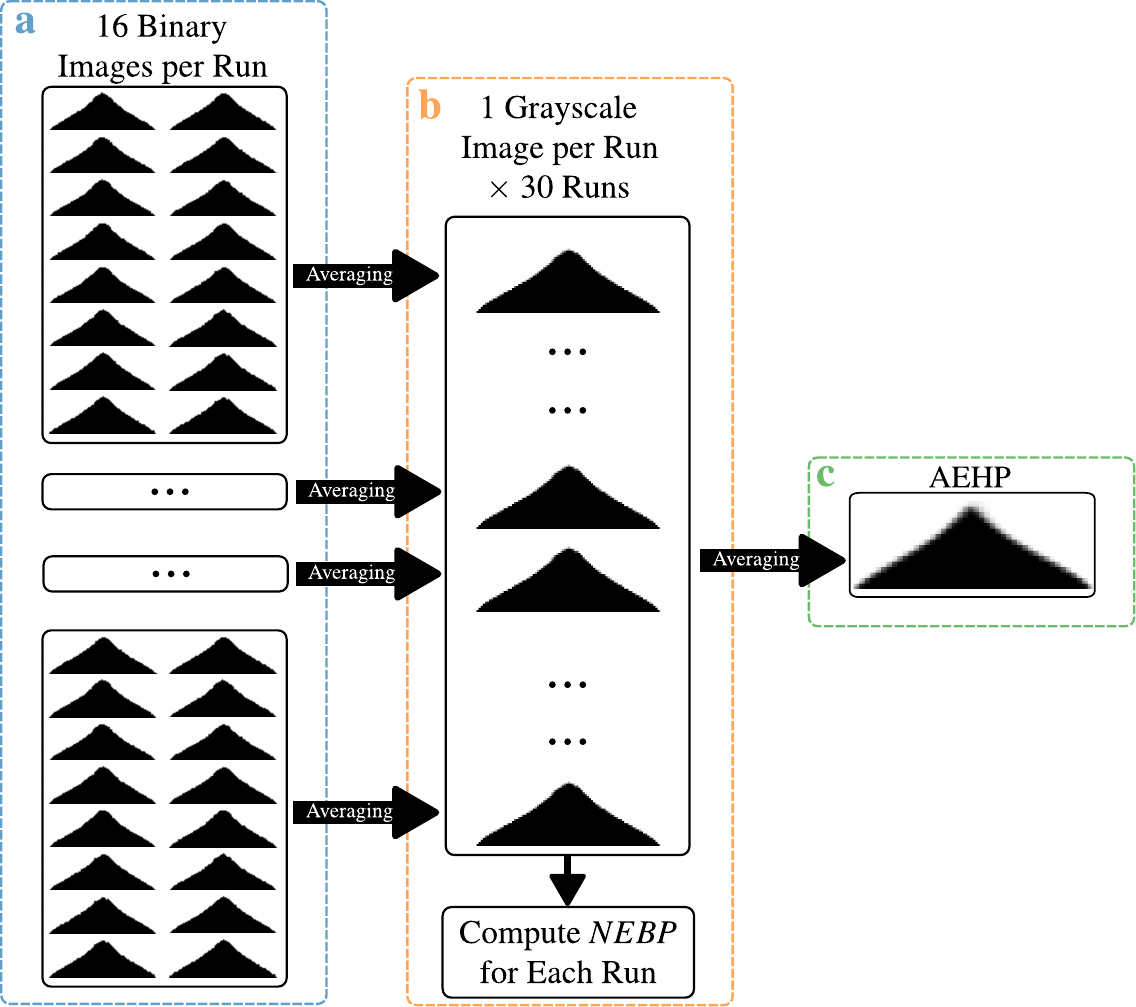}
        \caption{\textbf{Experimental methodology flow chart.} \textbf{(a)} 16 Binary images are taken for each experimental run perform on the GranuHeap. 30 runs are perform for a total of 480 binary images per preselected powder weight. \textbf{(b)} For every run, one grayscale image is computed by averaging the pixel values of the 16 associated binary images. The grayscale images are then post-process to compute their number of equivalent black pixel ($NEBP$) using \eqref{eq::NEBP}. Figure~\ref{fig::NEBP_Weight} presents the $NEBP$ values obtained for different preselected weights for each powder type. \textbf{(c)} Once the adequate powder weight is found, the average experimental heap profile (AEHP) used for the calibration process is computed by averaging the grayscale images associated with the 30 runs performed.}
        \label{fig::Experimental_flow_chart}
\end{figure}

\begin{equation}
    NEBP = \frac{1}{255}\sum_{p}I_p,
    \label{eq::NEBP}
\end{equation}
where $p$ are the pixels. Figures~\ref{fig::NEBP_Weight}a and b present the $NEBP$ obtained for six preselected weights for the titanium and aluminum powders, respectively. Note that the weights used for each powder corresponds to the same volume of raw material, for instance, $\SI[minimum-decimal-digits=2]{1.50}{\gram}$ of Ti6Al4V and $\SI{0.62}{\gram}$ of Al6061 both correspond to $\SI{0.227}{\centi\meter\cubed}$. Table~\ref{tab::powder_weight} shows the selected powder weights for each type, which are $\SI[minimum-decimal-digits=2]{1.50}{\gram}$ for the Ti6Al4V and $\SI{0.62}{\gram}$ for the Al6061. From this first step, the AEHP that will be used for the calibration procedure is created by averaging the 480 images taken from the 30 GranuHeap experiments associated with the optimal powder weight (Figure~\ref{fig::Experimental_flow_chart}c). 

\begin{table*}[!hbt]
    \caption{\textbf{Optimum powder weight.} Powder weight used in every DEM simulations for each powder type.}
    \centering
    \begin{tabular}{c | c}
        Powder type & Optimum powder weight  \\
        \hline
        Ti6Al4V & $\SI[minimum-decimal-digits=2]{1.50}{\gram}$ \\
        Al6061 & $\SI{0.62}{\gram}$ 
        \end{tabular}
        \label{tab::powder_weight}
\end{table*}

\section{Numerical Model}
\label{sec:numerical_model}
The DEM simulations performed in this work are carried out using \lethe{} \cite{alphonius2026, golshan2023} which is based on \dealii{}, a \texttt{C++} finite element library \cite{africa_2024} with an extensive particle module \cite{gassmoller2018}. This section briefly explains the DEM as well as the models used to compute the interaction forces and torques applied to particles. For conciseness, only the equations for particle–particle interactions are presented, as particle–wall models follow the same principles. Further details on these models can be found in previous works \cite{alphonius2026, gaboriault2026}.

The DEM is a numerical method used to simulate granular systems. It relies on solving Newton’s second law of motion \eqref{eq::dem_newton} and Euler’s law of angular motion \eqref{eq::dem_euler} numerically for every particle using an explicit time integration scheme, velocity Verlet in our case. To do so, interaction forces and torque are computed at each time step using a Lagrangian frame of reference.  
\begin{align}
     m_{i}\frac{d \vm{v}_{i}}{dt} &= m_{i}\vm{g} + \sum_{j \in \mathrm{C}_i}  \vm{F}_{\mathrm{n},ij} + \vm{F}_{\mathrm{t},ij} + \vm{F}_{\mathrm{c},ij}  \label{eq::dem_newton},\\
     I_{i}\frac{d\vm{\omega}_{i}}{dt} &= \sum_{j \in \mathrm{C}_i} \vm{M}_{\mathrm{t},i} +  \vm{M}_{\mathrm{r},ij},
     \label{eq::dem_euler}
\end{align}
where $i$ and $j$ are the particle indices. $j \in \mathrm{C}_i$ denotes all particles $j$ currently in interaction with particle $i$. $m_i$ is the particle mass and $I_i$ is the moment of inertia. $\vm{g}$ is the gravitational acceleration. $\vm{v}_i$ and $\vm{\omega}_i$ are the translational and angular velocities, respectively. $\vm{F}_{\text{n},ij}$ is the normal contact force, $\vm{F}_{\text{t},ij}$ is the tangential contact force and $\vm{F}_{\text{c},ij}$ is the cohesive force due to the Van der Waal interactions. Finally, $\vm{M}_{\text{t},ij}$ is the torque resulting from the tangential force and $\vm{M}_{\text{r},ij}$ is the rolling resistance torque used to model the non-sphericity of particles. The following subsections describe how those forces and torques are computed. 

\subsection{Contact force and torque}
\label{subsec:contact_force}
We use a non-linear visco-elastic model to compute $\vm{F}_{\mathrm{n},ij}$, $\vm{F}_{\mathrm{t},ij}$ from the overlap ($\overlap{}$) using \cref{eq::dem_normal_overlap}, the tangential displacement ($\tdisplacement{}$) from \cref{eq::dem_tangential_overlap} and the relative velocity ($\vm{v}_{ij}$) from \cref{eq::dem_tot_rel_vel}. Afterwards, $\vm{M}_{\mathrm{t},ij}$ is computed using $\vm{F}_{\mathrm{t},ij}$ in \cref{eq::tangential_torque}.

\begin{align}
    \delta_{\mathrm{n}} &= R_i + R_j - \norm{\vm{x}_j - \vm{x}_i} \label{eq::dem_normal_overlap} , \\
    \vm{n}_{ij} &= \frac{\vm{x}_j - \vm{x}_i}{\norm{\vm{x}_j - \vm{x}_i}},\\
    \vm{v}_{ij} &= \vm{v}_j - \vm{v}_i + \pp{R_i \vm{\omega}_i + R_j \vm{\omega}_j} \times \vm{n}_{ij} \label{eq::dem_tot_rel_vel},\\ 
    \vm{v}_{\mathrm{n},ij} &=  \pp{\vm{n}_{ij} \cdot \vm{v}_{ij}} \vm{n}_{ij}, \label{eq::dem_normal_velocity}
    \\\vm{v}_{\mathrm{t},ij} &=\vm{v}_{ij} - \vm{v}_{\mathrm{n},ij} ,\label{eq::dem_tangential_velocity}\\
    \tdisplacement &= \int_{t_0^\mathrm{c}}^{t}\vm{v}_{\mathrm{t},ij}\mathrm{d}t,
    \label{eq::dem_tangential_overlap}
\end{align}
where $R$ denotes the particle radius and $\vm{x}$ its position. $\vm{n}_{ij}$ is the normal contact vector, $\vm{v}_{ij}$ is the total relative velocity at the point of contact with its normal and tangential components denoted as $\vm{v}_{\mathrm{n},ij}$ and $\vm{v}_{\mathrm{t},ij}$, respectively. $t_0^\mathrm{c}$ denotes the time at which the contact between the particles has started. From these quantities, the stiffness and damping coefficients, listed in Table~\ref{model:table:DEMForces} and needed in \cref{eq::normal_force} and \eqref{eq::tangential_force}, are computed.
\begin{table*}[!htbp]
\caption{Equations for the DEM contact and cohesive force models}
\centering
\begin{tabular}{|l l|}
  \hline
    Parameter                     & Equation \\           
    \hline
    Normal stiffness              & $k_{\mathrm{n}}= \frac{4}{3}Y_{\mathrm{e}} \sqrt{R_{\mathrm{e}}\delta_{\mathrm{n}}}$ \\
    Tangential stiffness          & $k_{\mathrm{t}}= 8G_{\mathrm{e}} \sqrt{R_{\mathrm{e}} \delta_{\mathrm{n}}}$ \\
    Normal damping                & $\eta_{\mathrm{n}}= -2\sqrt{\frac{5}{6}}  \frac{\ln\pp{e_\mathrm{r}}}{\sqrt{\ln^2\pp{e_\mathrm{r}} + \pi^2}} \sqrt{\frac{2}{3}k_{\mathrm{n}} m_{\mathrm{e}}}$ \\
    Tangential damping            & $\eta_{\mathrm{t}}= -2\sqrt{\frac{5}{6}}  \frac{\ln\pp{e_\mathrm{r}}}{\sqrt{\ln^2\pp{e_\mathrm{r}} + \pi^2}} \sqrt{k_{\mathrm{t}} m_{\mathrm{e}}}$ \\
    Effective mass                & $\frac{1}{m_{\mathrm{e}}} = \frac{1}{m_i} + \frac{1}{m_j}$ \\
    Effective radius              & $\frac{1}{R_{\mathrm{e}}} = \frac{1}{R_{i}} + \frac{1}{R_j}$ \\
    Effective Young's modulus     & $\frac{1}{Y_{\mathrm{e}}} = \frac{\pp{1-\nu_i^2}}{Y_i} + \frac{\pp{1-\nu_j^2}}{Y_j}$ \\
	Effective shear modulus      & $\frac{1}{G_{\mathrm{e}}} = \frac{2\pp{2+\nu_i}\pp{1-\nu_i}}{Y_i} + \frac{2\pp{2+\nu_j}\pp{1-\nu_j}}{Y_j}$ \\
	Effective sliding friction coefficient                & $\mu_{\mathrm{t}} = \pp{\frac{1}{\mu_{\mathrm{t},i}} + \frac{1}{\mu_{\mathrm{t},j}}}^{-1}$ \\
	Effective rolling friction coefficient                &  $\mu_{r} = \pp{\frac{1}{ \mu_{r,i}}  + \frac{1}{ \mu_{r,j}}}^{-1}$\\
    Effective restitution coefficient     & $e_r = \pp{\frac{1}{ e_{r,i}}  + \frac{1}{ e_{r,j}}}^{-1}$ \\
    Effective surface energy  & $\gamma_{\mathrm{e}} = \gamma_{i} + \gamma_{j} -\pp{\sqrt{\gamma_{i}} - \sqrt{\gamma_{j}}}^{2}$ \\
    Poisson ratio of particle $i$  		        & $\nu_i$ \\
    \hline
    \end{tabular}
\label{model:table:DEMForces}
\end{table*}

\begin{align}
    \vm{F}_{\mathrm{n},ij} &= k_\mathrm{n} \delta_\mathrm{n} \vm{n}_{ij} + \eta_\mathrm{n} \vm{v}_{\mathrm{n},ij} \label{eq::normal_force}, \\
    \vm{F}_{\mathrm{t},ij} &= \min \pp{ k_\mathrm{t} \vm{\delta}_\mathrm{t} + \eta_\mathrm{t} \vm{v}_{\mathrm{t},ij}, \ \mu_\mathrm{t} \norm{\vm{F}_{\mathrm{n},ij}} \frac{k_\mathrm{t} \vm{\delta}_\mathrm{t} + \eta_\mathrm{t} \vm{v}_{\mathrm{t},ij}}{\norm{k_\mathrm{t} \vm{\delta}_\mathrm{t} + \eta_\mathrm{t} \vm{v}_{\mathrm{t},ij}}}} \label{eq::tangential_force}, \\
    \vm{M}_{\mathrm{t},ij} &= - R_i (\vm{F}_{\mathrm{t},ij} \times  \vm{n}_{ij} ). \label{eq::tangential_torque}
\end{align} 

\subsection{Cohesive force}
\label{subsec:conhesive_force}
Due to the small size of the powders used within this work, an adequate cohesive force model must be used \cite{gaboriault2026, coetzee_2023a, seville_2000}. Here, we used a version of the Derjaguin–Muller–Toporov (DMT) \cite{derjaguin_1975} model introduced by \citet{meier2019a}, which is also used in previous work \cite{gaboriault2026}. 
\begin{equation}
        \vm{F}_{\mathrm{c},ij} \cdot \vm{n}_{ij} =    
        \begin{cases}
            F_{\mathrm{po}}, & \delta_{\mathrm{o}}  \leq \delta_\mathrm{n} \\
            \dfrac{-AR_{\mathrm{e}}}{6 \delta_{n}^2}, & \delta^*  < \delta_\mathrm{n} < \delta_{\mathrm{o}} \\
            0, &  \delta_{\mathrm{n}} \leq \delta^*
        \end{cases}
        \label{eq::DMT}
\end{equation}
with,
\begin{align}
    F_{\mathrm{po}} &= -2\pi\gamma_{\mathrm{e}}R_{\mathrm{e}},\\
    \delta_{\mathrm{o}} &= - \sqrt{\frac{ A R_{\mathrm{e}}}{6 \norm{{F_{\mathrm{po}}}}}} \label{eq::delta_o},\\
    \delta^* &= \frac{\delta_{\mathrm{o}}}{ \sqrt{C_{\mathrm{FPO}}}},
    \label{eq::delta_star}
\end{align}
where $F_{\mathrm{po}}$ is the pull-off force between the particles, $A$ is the Hamaker constant associated with the two particles in contact and $C_{\mathrm{FPO}}$ is a user-defined parameter, set to 0.1 in this work, that controls the distance at which long-range cohesive forces are considered negligible relative to $F_{\mathrm{po}}$. 

\subsection{Rolling friction}
\label{subsec:rol_fric}
Choosing the wrong rolling friction model can have a drastic effect on powder rheology in a DEM simulations. Since we are interested in the profile of a static powder heap, we use the elastic–plastic spring–dashpot (EPSD) model, as other rolling friction models either do not reach full static equilibrium or fail to produce physically realistic powder heaps \cite{ai_2011}.
\begin{subequations}
\begin{align}
   \vm{\omega}_{ji} &= \vm{\omega}_{i} - \vm{\omega}_{j}, \\
   \vm{\omega}_{ji,\mathrm{p}} &= \vm{\omega}_{ji} - \pp{\vm{\omega}_{ji}\cdot\vm{n}_{ij}}\vm{n}_{ij},\\
   \vm{\Delta\theta} &= \Delta t \; \vm{\omega}_{ji,\mathrm{p}},\\
   k_\mathrm{r} &= 2.25 k_\mathrm{n} \pp{\muRE \Ree}^2,\\
   \vm{\Delta M}_{\mathrm{r},t} ^\mathrm{k} &= - k_\mathrm{r}\vm{\Delta\theta},\\
   \vm{M}_{\mathrm{r}, t} ^\mathrm{k} &=  \vm{M}_{\mathrm{r},t-\Delta t}^\mathrm{k} + \vm{\Delta M}_{\mathrm{r},t}^\mathrm{k}, \\
   M^\mathrm{m}_\mathrm{r} &= \muRE R_e \norm{\vm{F}_{n,ij}},\\
   \vm{M}^k_{\mathrm{r},t} &= \begin{cases}
         \vm{M}^\mathrm{k}_{\mathrm{r},t}, & \norm{\vm{M}^\mathrm{k}_{\mathrm{r},t}} < M^\mathrm{m}_\mathrm{r} \\
         \frac{\vm{M}^\mathrm{k}_{\mathrm{r},t} }{\norm{\vm{M}^\mathrm{k}_{\mathrm{r},t}}} M^\mathrm{m}_\mathrm{r}, & \mathrm{else}  
    \end{cases},\\
    C_r &= 2 \eta_r \sqrt{I_e K_r},\\
    \vm{M}^d_{r,t} &= \begin{cases}
         -C_r \vm{\omega}_{ij,t} , & \norm{\vm{M}^k_{t}} < M^m_r \\
         -f C_r \vm{\omega}_{ij,t}, & \mathrm{else} 
    \end{cases},\\
    \vm{M}_{\mathrm{r},ij} &= \vm{M}_{\mathrm{r},t} = \vm{M}^k_{\mathrm{r},t} + \vm{M}^d_{\mathrm{r},t},
\end{align}\label{dem:rol_fric:EPSD}
\end{subequations}
where $\vm{\omega}_{ji}$ is the relative angular velocity between the particles $i$ and $j$, $\vm{\omega}_{ij,p}$ is the relative angular velocity in the contact plane, $\vm{\Delta\theta}$ is the incremental relative rotation, $k_r$ is the rolling stiffness constant, $\vm{\Delta M}_{r,t}$ is the incremental elastic rolling resistance torque, $M^\mathrm{m}_\mathrm{r}$ is the limiting spring torque norm, $C_r$ is the rolling viscous damping constant, and $f$ is the full mobilization model parameter, set to $0.1$ in this study. $\vm{M}^k_{\mathrm{r},t}$ and $\vm{M}^d_{\mathrm{r},t}$ are the elastic and viscous damping torques, respectively.

\section{Calibration metrics}
\label{sec::calibration_methodology}
The angle of repose (AOR) is commonly used to calibrate DEM models \cite{li2020a}. Although its definition varies in the literature depending on the application \cite{beakawial-hashemi2018}, it can generally be described as the angle formed between the free surface of a granular material and the horizontal plane. Its widespread use is mainly due to its simplicity, as it can be measured using various experimental techniques \cite{li2020a}.

\begin{figure}[!htb]
        \centering
        \includegraphics[width=70mm]{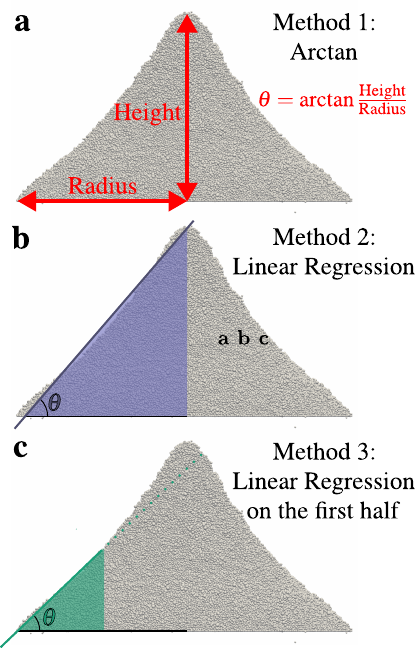}
        \caption{\textbf{Angle of Repose (AOR) measurement methods:} \textbf{(a)} First method that relies on the ratio between the height of the powder heap and the radius at its base. \textbf{(b)} Second method using a linear regression using the entire slope of one side of the heap. \textbf{(c)} Third method using a linear regression across a fraction of the slope of one side of the heap, half of it in this case. For this specific example, these methods give $\SI{48.65}{\degree}$, $\SI{49.34}{\degree}$ and $\SI{41.50}{\degree}$, respectively.}
        \label{fig::AOR_method}
\end{figure}

However, this apparent simplicity also introduces challenges, particularly when dealing with cohesive powders, which are the focus of this work. Figure~\ref{fig::AOR_method} illustrates different methods that can be used to measure the AOR of a cohesive powder heap. The first method is based on the ratio between the height of the heap and the radius of its base \cite{beakawial-hashemi2018} (see Figure~\ref{fig::AOR_method}a). The second and third methods both rely on a linear regression of the slope formed by the powder heap  \cite{muller2021a, katagiri2026}, but differ in the portion of the slope considered. The second methods uses the entire slope for the linear regression, while the third method uses only the first half starting from the base (see Figures~\ref{fig::AOR_method}b-c). Depending on the method used, the measured angle is not the same. This is not a problem in itself, provided that the same method is consistently applied to both the experiments and the simulations. The main difficulty arises because many different parameter combinations can yield the same AOR for a given measurement method. In other words, the DEM parameter solution space that offers the same AOR as the experimental run is large, or is non-unique, for a given AOR measurement method. One possible explanation for this is that the AOR does not give enough information about the powder behaviour, especially when cohesion is non negligible. Thus, it is highly probable that the calibration of a given cohesive powder using different AOR measurement methods, or even the same method, can lead to significantly different parameter sets.

A second challenge arises from the heap morphology formed by cohesive powders. The three methods illustrated in Figure~\ref{fig::AOR_method} implicitly assume an idealized conical heap, characterized by perfect axisymmetry and a linear free surface profile. However, in the case of cohesive powders, these assumptions are often invalid. Consequently, the reliance on a single geometric parameter, like the AOR, to characterize the powder heaps is not reliable.

\subsection{Proposed calibration metric based on image analysis}
\label{sec::sad_metric}
We propose a calibration metric based on the comparison of heap profiles using their pixel values. The first image is the AEHP (Figure~\ref{fig::Experimental_flow_chart}c), obtained from the optimum powder weight selected in Table~\ref{tab::powder_weight}. The grayscale image associated with a given DEM calibration simulation, referred to as the average numerical heap profile (ANHP), is generated using an in-house \paraview{} script that takes 16 binary images taken over a $\SI{180}{\degree}$ arc around the simulated static heap and averages them. The sum of absolute differences (SAD) between the AEHP and the ANHP can then be computed:
\begin{equation}
    \mathrm{SAD} = \frac{1}{255}\sum_p|\mathrm{AEHP}_p - \mathrm{ANHP}_p|.
\end{equation}
The DEM simulation with the lowest SAD is considered to provide the best agreement with the calibration experiment. 

\section{Results} 
\label{sec::results}
In this section, we demonstrate the efficacy of the proposed metric by calibrating the powders introduced in Section~\ref{subsec:powders}. The calibration is performed by varying the friction coefficient ($\mu_{t,i}$), the rolling friction coefficient ($\mu_{r,i}$), and the surface energy ($\gamma_i$), as the other parameters have been judged less sensitive. Table~\ref{tab::dem_model_param} presents the remaining DEM model parameters and powder physical properties used in the simulations. The Young's modulus of the Ti6Al4V powder is kept at $\SI{26.25}{\mega\pascal}$ to remain consistent with previous work \cite{gaboriault2026}, while a lower value of $\SI{10.00}{\mega\pascal}$ is used for the Al6061 powder to reduce the computational cost of each calibration simulation. The time step used for the Ti6Al4V and Al6061 powders is $2.816 \times 10^{-7}$~\si{\second} and $2.041 \times 10^{-7}$~\si{\second}, respectively, corresponding to $15\%$ of the Rayleigh time of the smallest particle used in the simulation of each powder type.
\begin{table*}[!hbp]
    \caption{\textbf{DEM model parameter and physical properties used in the calibration simulations.}}    \centering
    \begin{tabular}{c | c | c}
        Parameter & Ti6Al4V & Al6061 \\
        \hline
        Young's modulus ($Y_i$) [\si{\mega\pascal}]             & 26.25 & 10.00 \\
        Poisson's Ratio ($\nu_i$) [-]                           & 0.342 & 0.33 \\
        True density ($\rho$) [\si{\kilo\gram\per\meter\cubed}] & 4386  & 2700 \\
        Restitution coefficient ($e_i$) [-]                     & 0.9   & 0.9 \\
        Rolling viscous damping ($\eta_\mathrm{r}$)             & 0.3   & 0.3 \\
        $f$ [-]                                                 & 0.1   & 0.1\\
    \end{tabular}
    \label{tab::dem_model_param}
\end{table*}

\begin{figure}[!t]
        \centering
        \includegraphics[width=90mm]{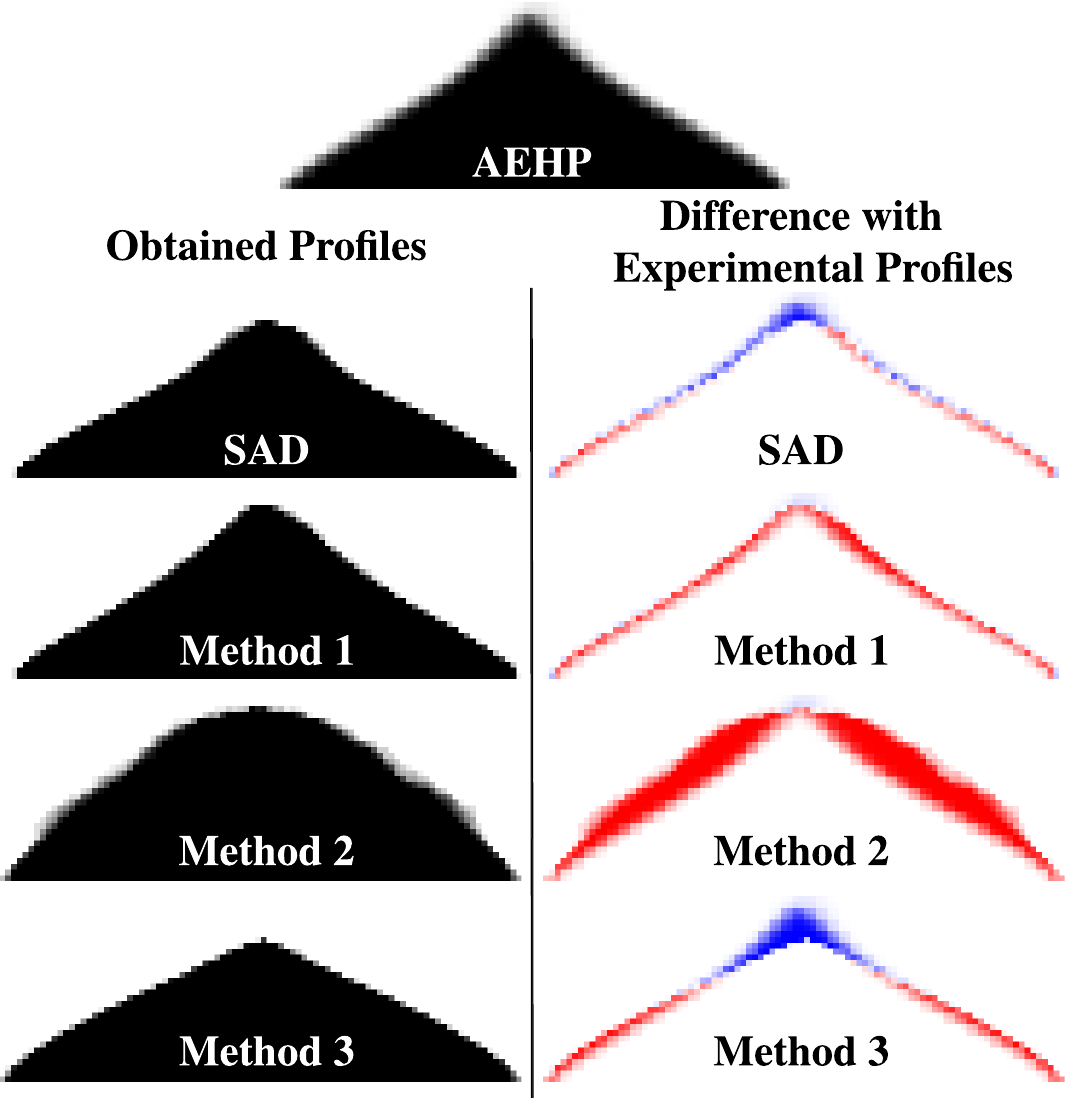}
        \caption{\textbf{Comparison of the powder heap profiles obtained using each calibration metric for the Ti6Al4V powder.} The top profile represents the Average experimental heap profile (AEHP). Each subsequent row corresponds to the profile with the best agreement with a specific calibration metric (SAD, Method 1, Method 2, and Method 3). The left column shows the average numerical heap profile (ANHP) obtained for the associated parameter set listed in Table~\ref{tab::best_calib_per_method_Ti}. The right column illustrates the signed error between the simulated and experimental profiles using \eqref{eq::signed_error}. A blue pixel represent positive values (underprediction) and red pixels represent negative values (overprediction). White pixels indicate perfect agreement between the numerical and experimental average profiles.}
        \label{fig::comparaison_method_Ti}
\end{figure}

\begin{figure}[!htb]
        \centering
        \includegraphics[width=90mm]{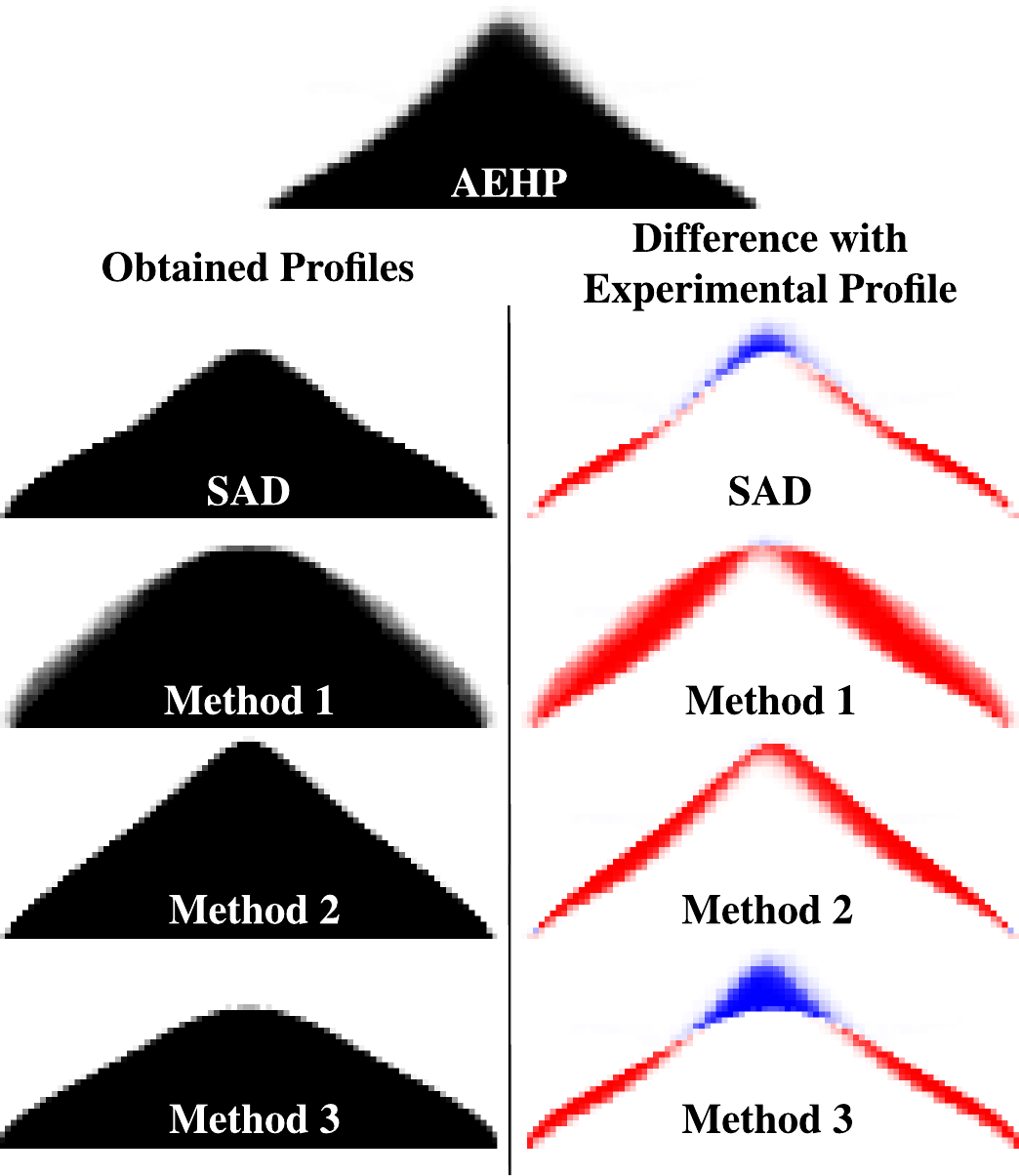}
        \caption{\textbf{Comparison of the powder heap profiles obtained using each calibration metric for the Al6061 powder.} The top profile represents the Average experimental heap profile (AEHP). Each subsequent row corresponds to the profile with the best agreement with a specific calibration metric (SAD, Method 1, Method 2, and Method 3). The left column shows the average numerical heap profile (ANHP) obtained for the associated parameter set listed in Table~\ref{tab::best_calib_per_method_Ti}. The right column illustrates the signed error between the simulated and experimental profiles using \eqref{eq::signed_error}. A blue pixel represent positive values (underprediction) and red pixels represent negative values (overprediction). White pixels indicate perfect agreement between the numerical and experimental average profiles.}
        \label{fig::comparaison_method_Al}
\end{figure}

For Ti4Al4V powder, a full-factorial parametric sweep is first performed using values of $0.001$, $0.1$, $0.2$, $0.3$, and so on up to $0.7$ for $\mu_{t,i}$ and $\mu_{r,i}$, respectively. The values considered for $\gamma_i$ are $1 \times 10^{-6}$, $5 \times 10^{-5}$, $10 \times 10^{-5}$, $15 \times 10^{-5}$, and so on up to $40 \times 10^{-5}$. A second round of simulations is then performed by selecting the 20 simulations with the lowest SAD values and sampling around them using half-step increments, thus creating 26 new sampling points per selected simulations. In an effort to reduce computational cost, a Latin Hypercube Sampling (LHS) is performed for the Al6061 powder instead due to its smaller particle size distribution (PSD), which requires simulating 3.75 times more particles. The minimum and maximum values of the sampled domain are kept identical to those used for the Ti6Al4V powder. In total, 670 and 100 simulations are launched for the Ti6Al6V and Al6061 powders, respectively. Tables~\ref{fig::comparaison_method_Ti}
and \ref{fig::comparaison_method_Al} list the optimum parameter sets for each calibration method for those same two powders.

\begin{table*}[!htbp]
    \caption{\textbf{Best parameter sets obtained and corresponding measured quantity for each calibration method for the Ti6Al4V powder.} For the SAD method, the measured quantity is reported as Equivalent Black Pixels (EBP), while for Methods 1--3 it corresponds to the measured angle of repose (AOR). The reference angles for Methods 1--3 are $\SI{34.90}{\degree}$, $\SI{33.73}{\degree}$ and $\SI{31.34}{\degree}$, respectively}
    \centering
    \begin{tabular}{c | c | c | c | c}
        Calibrated parameters & SAD & Method 1 & Method 2 & Method 3 \\
        \hline
        $\mu_{t,i} [-]$                              & $0.25$            & $0.35$                & $0.15$                & $0.60$    \\
        $\mu_{r,i} [-]$                              & $0.001$           & $0.001$               & $0.30$                & $0.10$   \\
        $\gamma_i [\SI{}{\joule\per\meter\squared}]$ & $40\times10^{-5}$ & $65\times 10^{-5}$    & $35\times 10^{-5}$    & $5\times 10^{-5}$  \\
        Measured quantity                            & $82.72$ EBP       & $\SI{34.90}{\degree}$ & $\SI{33.61}{\degree}$ & $\SI{40.56}{\degree}$  \\
    \end{tabular}
    \label{tab::best_calib_per_method_Ti}
\end{table*}
\begin{table*}[!htbp]
    \caption{\textbf{Best parameter sets obtained and corresponding measured quantity for each calibration method for the Al6061 powder.} For the SAD method, the measured quantity is reported as Equivalent Black Pixels (EBP), while for Methods 1--3 it corresponds to the measured angle of repose (AOR). The reference angles for Methods 1--3 are $\SI{37.30}{\degree}$, $\SI{38.43}{\degree}$ and $\SI{30.69}{\degree}$, respectively}  
    \centering
    \begin{tabular}{c | c | c | c | c}
        Calibrated parameters & SAD & Method 1 & Method 2 & Method 3 \\
        \hline
        $\mu_{t,i} [-]$                              & $0.6819$             & $0.2656$               & $0.5714$                & $0.2331$    \\
        $\mu_{r,i} [-]$                              & $0.09323$            & $0.0635$               & $0.5523$                & $0.0729$   \\
        $\gamma_i [\SI{}{\joule\per\meter\squared}]$ & $5.843\times10^{-5}$ & $19.160\times 10^{-5}$ & $2.101\times 10^{-5}$ & $7.920\times 10^{-5}$  \\
        Measured quantity                            & $164.63$ EBP         & $\SI{36.10}{\degree}$  & $\SI{38.38}{\degree}$   & $\SI{31.02}{\degree}$\\
    \end{tabular}
    \label{tab::best_calib_per_method_Al}
\end{table*}

Using the AEHP and the ANHPs obtained from the calibration metrics, the local differences between the numerical and experimental heap profiles are visualized through the signed difference map (SDM), defined as

\begin{equation}
\mathrm{SDM}_p = \mathrm{ANHP}_p - \mathrm{AEHP}_p,
\label{eq::signed_error}
\end{equation}

where each pixel represents the local excess or deficit of material in the average numerical heap profile relative to the experimental one. Positive values indicate an underprediction of the heap profile, while negative values indicate an overprediction. Figure~\ref{fig::comparaison_method_Ti} presents the AEHP together with the best ANHPs obtained for each calibration metric for the Ti6Al4V powder. The right column of the same figure shows the corresponding SDM for each calibration metric, where blue pixels represent positive values (underprediction) and red pixels represent negative values (overprediction). White pixels indicate perfect agreement between the numerical and experimental average profiles. Good agreement between the ANHP and AEHP is observed when using the SAD and Method 1, as indicated by the small number of colored pixels and their low intensity, while poor agreement is observed for the Method 2 and 3. This proves that AOR based measurements methods do not guarantee finding a DEM model parameter set that adequately reproduce the real powder behaviour. 

Figure~\ref{fig::comparaison_method_Al} present the same results as Figure~\ref{fig::comparaison_method_Ti}, but for the Al6061 powder. The profile obtained from the SAD metric is able to reproduce the concave slope also present in the AEHP while the AOR measurement metrics miss this key feature. The best profile obtained with Method 1 present a convex slope and overpredicts the profile everywhere along the width of the heap. Method 2 does the same thing to a lesser extent. Method 3 has a concave slope, but lacks the pointy tip of the heap. 


\begin{figure}[!htb]
        \centering
        \includegraphics[width=70mm]{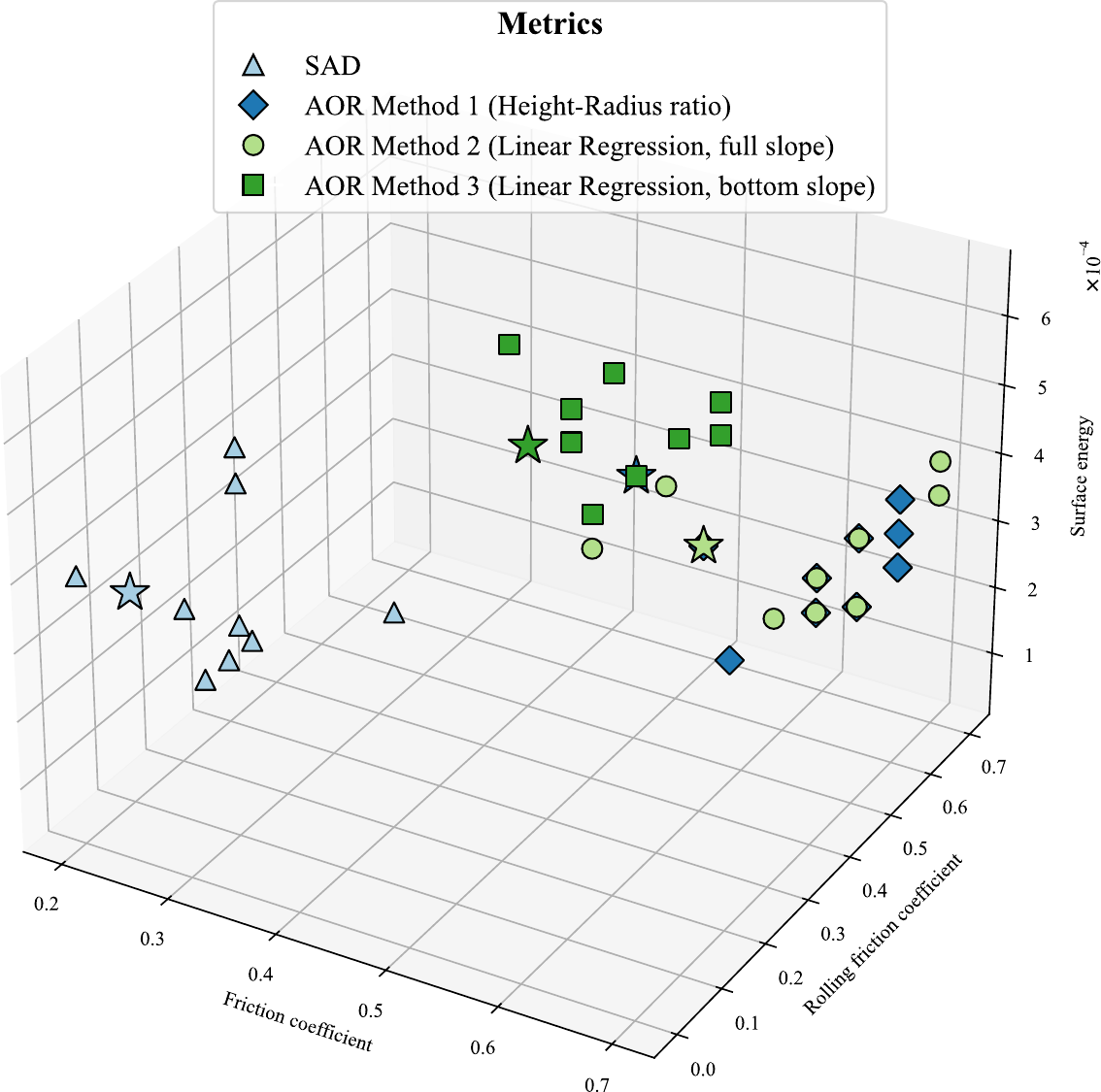}
        \caption{\textbf{Top 10 propertie sets for each calibration metric for the Ti6Al4V powder.} Each color-marker combination corresponds to a different calibration metric, while the star marker indicates the best-performing parameter set for that metric. Distinct clusters highlight that different metrics favor different combinations of DEM parameters.}
        \label{fig::scatter_plot_Ti}
\end{figure}

Figure~\ref{fig::scatter_plot_Ti} presents, for each calibration metric, the top ten data points simulated using the Ti6Al4V powder. The best result for each metric is identified with a star marker. Distinct data clusters are clearly visible, indicating that different metrics emphasize different physical mechanisms governing the powder heap formation. For the SAD metric, the cluster is characterized by low $\mu_{t,i}$, low $\mu_{r,i}$, and high $\gamma_i$. In contrast, the cluster associated with AOR Method 1 (Height-Radius ratio) lies at the opposite end of the parameter space, with high $\mu_{t,i}$, high $\mu_{r,i}$, and low $\gamma_i$.

\section{Conclusion} 
In this work, a novel calibration metric based on the comparison between experimental and numerical average heap profiles is introduced. This metric relies on the pixel-wise intensity differences between the corresponding grayscale images of both profiles. This metric shows promising capabilities by its ability to better capture the heap morphology compared to traditional AOR based calibration metrics. 

The results obtained for both Ti6Al4V and Al6061 powders demonstrate that traditional AOR-based calibration methods can lead to significantly different and non-unique sets of DEM parameters, even when reproducing similar heap angles. This highlights the intrinsic limitations of scalar metrics in capturing the complex behaviour of cohesive granular materials. In contrast, the proposed image-based metric is able to distinguish between different heap morphologies and provides improved sensitivity to local features such as slope curvature and peak sharpness, which can be neglected by AOR based approaches.

The analysis of the calibrated parameter spaces further reveals that different calibration metrics promote distinct regions of the DEM parameter space, confirming that the choice of calibration method leads to different predicted behaviour of the simulated powder. In particular, the proposed SAD metric leads to a more consistent agreement between experimental and numerical heap profiles, as confirmed by both qualitative comparisons and signed difference maps. This suggests that the SAD metric leads to calibrated model parameter sets that better reproduce the powder behaviour in the target simulations for which the calibration is intended.

Overall, this study demonstrates that image-based calibration metrics offer a more robust and informative alternative to classical scalar approaches for DEM model calibration of cohesive powders. Future work will focus on assessing the influence of the best model parameter sets obtained using the different metrics on real world processes with experimental validation.   

\section{Acknowledgements}

 The authors acknowledge technical support and computing time provided by the Digital Research Alliance of Canada and Calcul Québec. The authors would like to thank Prof. Jean-Philippe Harvey and Semplor for providing the SEM images of the Ti6Al4V powder. BB and DM acknowledges financial support from the National Research Council (NRC) Canada through the Collaborative R\&D Initiative grant AM-137-1. BB acknowledges the funding from the Multiphysics Multiphase Intensification Automatization Workbench (MMIAOW) Canadian Research Chair Level 2 in computer-assisted design and scale-up of alternative energy vectors for sustainable chemical processes (CRC-2022-00340).

\section*{CRediT authorship contribution statement}
\textbf{Olivier Gaboriault: }Writing – original draft, Conceptualization, Data curation, Formal analysis, Investigation, Methodology (Numerical), Project administration, Software, Visualization.
\textbf{Antonella Succar: }Writing – original draft, Conceptualization, Data curation, Formal analysis, Investigation, Methodology (Numerical), Software, Visualization.
\textbf{Cléo Delêtre: }Writing – original draft, Conceptualization, Data curation, Formal analysis, Investigation, Methodology (Numerical), Software.
\textbf{Anatolie Timercan: }Writing – review \& editing, Investigation, Methodology (Experimental).
\textbf{Roger Pelletier: }Writing – review \& editing, Funding acquisition, Methodology (Experimental), Resources, Supervision.
\textbf{David Melancon: }Writing – review \& editing, Conceptualization, Data curation, Funding acquisition, Methodology (Numerical), Resources, Supervision, Visualization.
\textbf{Bruno Blais: }Writing – review \& editing, Conceptualization, Data curation, Funding acquisition, Methodology (Numerical), Resources, Software, Supervision, Visualization.

\section*{Declaration of Competing Interest}
The authors declare that they have no known competing financial interests or personal relationships that could have appeared to influence the work reported in this paper.
\appendix

\section*{Data availability}
Data will be made available on request.

\bibliographystyle{unsrtnat}
\bibliography{bibliographie}
\label{app1}

\end{document}

%% file: commands.tex
\newcommand{\pp}[1]{\left(#1\right)}
\newcommand{\norm}[1]{\left|#1\right|}

\newcommand{\vm}[1]{\bm{#1}}

\newcommand{\dealii}[0]{\texttt{deal.II}}
\newcommand{\lethe}[0]{\texttt{Lethe}}
\newcommand{\paraview}[0]{\texttt{Paraview}}

\newcommand*\justify{%
  \fontdimen2\font=0.4em
  \fontdimen3\font=0.2em
  \fontdimen4\font=0.1em
  \fontdimen7\font=0.1em
  \hyphenchar\font=`\-
}

\renewcommand{\texttt}[1]{%
  \begingroup
  \ttfamily
  \begingroup\lccode`~=`/\lowercase{\endgroup\def~}{/\discretionary{}{}{}}%
  \begingroup\lccode`~=`[\lowercase{\endgroup\def~}{[\discretionary{}{}{}}%
  \begingroup\lccode`~=`.\lowercase{\endgroup\def~}{.\discretionary{}{}{}}%
  \catcode`/=\active\catcode`[=\active\catcode`.=\active
  \justify\scantokens{#1\noexpand}%
  \endgroup
}

%% file: elsarticle-template-num.bbl
\begin{thebibliography}{29}
\providecommand{\natexlab}[1]{#1}
\providecommand{\url}[1]{\texttt{#1}}
\expandafter\ifx\csname urlstyle\endcsname\relax
  \providecommand{\doi}[1]{doi: #1}\else
  \providecommand{\doi}{doi: \begingroup \urlstyle{rm}\Url}\fi

\bibitem[Sames et~al.(2016)Sames, List, Pannala, Dehoff, and Et.]{sames2016}
W.~J. Sames, F.~A. List, S.~Pannala, R.~R. Dehoff, and Al~Et.
\newblock The metallurgy and processing science of metal additive manufacturing.
\newblock \emph{International Materials Reviews}, 2016.
\newblock \doi{10.1080/09506608.2015.1116649}.

\bibitem[Jadidi et~al.(2025)Jadidi, Ebrahimi, {Ein-Mozaffari}, and Lohi]{jadidi2025}
Behrooz Jadidi, Mohammadreza Ebrahimi, Farhad {Ein-Mozaffari}, and Ali Lohi.
\newblock Calibration of {{DEM}} input parameters for simulation of the cohesive materials: {{Comparison}} of response surface method and machine learning models.
\newblock \emph{Particuology}, 100:\penalty0 214--231, May 2025.
\newblock ISSN 1674-2001.
\newblock \doi{10.1016/j.partic.2025.03.018}.

\bibitem[Cundall and Strack(1979)]{cundall_1979}
P.~A. Cundall and O.~D.~L. Strack.
\newblock A discrete numerical model for granular assemblies.
\newblock \emph{G\'eotechnique}, 29\penalty0 (1):\penalty0 47--65, March 1979.
\newblock ISSN 0016-8505.
\newblock \doi{10.1680/geot.1979.29.1.47}.

\bibitem[Yim et~al.(2022)Yim, Bian, Aoyagi, Yamanaka, and Chiba]{yim2022}
Seungkyun Yim, Huakang Bian, Kenta Aoyagi, Kenta Yamanaka, and Akihiko Chiba.
\newblock Spreading behavior of {{Ti48Al2Cr2Nb}} powders in powder bed fusion additive manufacturing process: {{Experimental}} and discrete element method study.
\newblock \emph{Additive Manufacturing}, 49:\penalty0 102489, January 2022.
\newblock ISSN 2214-8604.
\newblock \doi{10.1016/j.addma.2021.102489}.

\bibitem[Yim et~al.(2025)Yim, Wang, Aoyagi, Yamanaka, and Chiba]{yim2025a}
Seungkyun Yim, Hao Wang, Kenta Aoyagi, Kenta Yamanaka, and Akihiko Chiba.
\newblock Comparative evaluation of powder spreading strategies to enhance powder bed quality in powder bed fusion additive manufacturing: {{A DEM}} simulation study.
\newblock \emph{Powder Technology}, 453:\penalty0 120614, March 2025.
\newblock ISSN 0032-5910.
\newblock \doi{10.1016/j.powtec.2025.120614}.

\bibitem[Wu et~al.(2026)Wu, Zou, Kanishka, Chiu, Wu, Qiao, Shen, An, and Huang]{wu2026}
Qiong Wu, Yi~Zou, Surendra~Archaryagie Kanishka, Louis Ngai~Sum Chiu, Yuhang Wu, Chuang Qiao, Haopeng Shen, Xizhong An, and Aijun Huang.
\newblock Challenges of powder spreading in laser powder bed fusion additive manufacturing of lattice structures: {{The}} phenomena, mechanisms, and solutions.
\newblock \emph{Journal of Manufacturing Processes}, 157:\penalty0 575--592, January 2026.
\newblock ISSN 1526-6125.
\newblock \doi{10.1016/j.jmapro.2025.12.016}.

\bibitem[Gaboriault et~al.(2026)Gaboriault, Timercan, Pelletier, Lefebvre, Melancon, and Blais]{gaboriault2026}
Olivier Gaboriault, Anatolie Timercan, Roger Pelletier, Louis-Philippe Lefebvre, David Melancon, and Bruno Blais.
\newblock Increase in packing density during multi-layer powder spreading: {{An}} experimental and numerical study.
\newblock \emph{Powder Technology}, 478:\penalty0 122460, July 2026.
\newblock ISSN 0032-5910.
\newblock \doi{10.1016/j.powtec.2026.122460}.

\bibitem[Herman et~al.(2022)Herman, Gan, Zhou, and Yu]{herman2022}
Angga~Pratama Herman, Jieqing Gan, Zongyan Zhou, and Aibing Yu.
\newblock Discrete particle simulation for mixing of granular materials in ribbon mixers: {{A}} scale-up study.
\newblock \emph{Powder Technology}, 400:\penalty0 117222, March 2022.
\newblock ISSN 0032-5910.
\newblock \doi{10.1016/j.powtec.2022.117222}.

\bibitem[Ferreira et~al.(2023)Ferreira, Geitani, Silva, Blais, and Lopes]{ferreira2023}
Victor~O. Ferreira, Toni~El Geitani, Daniel Silva, Bruno Blais, and Gabriela~C. Lopes.
\newblock In-depth validation of unresolved {{CFD-DEM}} simulations of liquid fluidized beds.
\newblock \emph{Powder Technology}, 426:\penalty0 118652, August 2023.
\newblock ISSN 0032-5910.
\newblock \doi{10.1016/j.powtec.2023.118652}.

\bibitem[Chen et~al.(2016)Chen, Wang, and Li]{chen2016}
Xizhong Chen, Junwu Wang, and Jinghai Li.
\newblock Multiscale modeling of rapid granular flow with a hybrid discrete-continuum method.
\newblock \emph{Powder Technology}, 304:\penalty0 177--185, December 2016.
\newblock ISSN 0032-5910.
\newblock \doi{10.1016/j.powtec.2016.08.017}.

\bibitem[Tian et~al.(2018)Tian, Su, Zhan, Geng, Xu, and Liu]{tian2018}
Tian Tian, Jinglin Su, Jinhui Zhan, Shujun Geng, Guangwen Xu, and Xiaoxing Liu.
\newblock Discrete and continuum modeling of granular flow in silo discharge.
\newblock \emph{Particuology}, 36:\penalty0 127--138, February 2018.
\newblock ISSN 1674-2001.
\newblock \doi{10.1016/j.partic.2017.04.001}.

\bibitem[Golshan et~al.(2023)Golshan, Munch, Gassm{\"o}ller, Kronbichler, and Blais]{golshan2023}
Shahab Golshan, Peter Munch, Rene Gassm{\"o}ller, Martin Kronbichler, and Bruno Blais.
\newblock Lethe-{{DEM}}: An open-source parallel discrete element solver with load balancing.
\newblock \emph{Computational Particle Mechanics}, 10\penalty0 (1):\penalty0 77--96, February 2023.
\newblock ISSN 2196-4378, 2196-4386.
\newblock \doi{10.1007/s40571-022-00478-6}.

\bibitem[Werner et~al.(2025)Werner, Nicu{\c s}an, Shaw, Seville, Jenkins, Ingram, and {Windows-Yule}]{werner2025}
D.~Werner, A.~L. Nicu{\c s}an, L.~Shaw, J.~P.~K. Seville, B.~D. Jenkins, A.~Ingram, and C.~R.~K. {Windows-Yule}.
\newblock Open-source {{DEM}} digital models of widely-used powder characterisation tools, {{Part I}}: {{Shear}} testing and powder rheology.
\newblock \emph{Powder Technology}, 466:\penalty0 121363, December 2025.
\newblock ISSN 0032-5910.
\newblock \doi{10.1016/j.powtec.2025.121363}.

\bibitem[Meier et~al.(2019)Meier, Weissbach, Weinberg, Wall, and John~Hart]{meier2019a}
Christoph Meier, Reimar Weissbach, Johannes Weinberg, Wolfgang~A. Wall, and A.~John~Hart.
\newblock Modeling and characterization of cohesion in fine metal powders with a focus on additive manufacturing process simulations.
\newblock \emph{Powder Technology}, 343:\penalty0 855--866, February 2019.
\newblock ISSN 0032-5910.
\newblock \doi{10.1016/j.powtec.2018.11.072}.

\bibitem[Forgber et~al.(2022)Forgber, Khinast, and Fink]{forgber2022}
T.~Forgber, J.~G. Khinast, and E.~Fink.
\newblock A hybrid workflow for investigating wide {{DEM}} parameter spaces.
\newblock \emph{Powder Technology}, 404:\penalty0 117440, May 2022.
\newblock ISSN 0032-5910.
\newblock \doi{10.1016/j.powtec.2022.117440}.

\bibitem[Yablokova et~al.(2015)Yablokova, Speirs, Van~Humbeeck, Kruth, Schrooten, Cloots, Boschini, Lumay, and Luyten]{yablokova2015}
G.~Yablokova, M.~Speirs, J.~Van~Humbeeck, J.~P. Kruth, J.~Schrooten, R.~Cloots, F.~Boschini, G.~Lumay, and J.~Luyten.
\newblock Rheological behavior of {$\beta$}-{{Ti}} and {{NiTi}} powders produced by atomization for {{SLM}} production of open porous orthopedic implants.
\newblock \emph{Powder Technology}, 283:\penalty0 199--209, October 2015.
\newblock ISSN 0032-5910.
\newblock \doi{10.1016/j.powtec.2015.05.015}.

\bibitem[Ashrafizadeh et~al.(2026)Ashrafizadeh, Ejtehadi, Pelletier, {Habibnejad-Korayem}, Haeri, and Yue]{ashrafizadeh2026}
Seyed~Masoud Ashrafizadeh, Omid Ejtehadi, Roger Pelletier, Mahdi {Habibnejad-Korayem}, Sina Haeri, and Stephen Yue.
\newblock Experimental and numerical investigation of re-coating process parameters on the spreadability of plasma-atomized powder.
\newblock \emph{Powder Technology}, 469:\penalty0 121851, February 2026.
\newblock ISSN 0032-5910.
\newblock \doi{10.1016/j.powtec.2025.121851}.

\bibitem[Cordova et~al.(2020)Cordova, Bor, {de Smit}, Campos, and Tinga]{cordova2020}
Laura Cordova, Ton Bor, Marc {de Smit}, M{\'o}nica Campos, and Tiedo Tinga.
\newblock Measuring the spreadability of pre-treated and moisturized powders for laser powder bed fusion.
\newblock \emph{Additive Manufacturing}, 32:\penalty0 101082, March 2020.
\newblock ISSN 2214-8604.
\newblock \doi{10.1016/j.addma.2020.101082}.

\bibitem[Alphonius et~al.(2026)Alphonius, Barbeau, Blais, Gaboriault, Gu{\'e}vremont, Lamouche, Laurentin, Marquis, Munch, Ferreira, {Papillon-Laroche}, Patience, Prieto~Saavedra, and Vaillant]{alphonius2026}
Amishga Alphonius, Lucka Barbeau, Bruno Blais, Olivier Gaboriault, Olivier Gu{\'e}vremont, Justin Lamouche, Pierre Laurentin, Oreste Marquis, Peter Munch, Victor~Oliveira Ferreira, H{\'e}l{\`e}ne {Papillon-Laroche}, Paul~Alexander Patience, Laura Prieto~Saavedra, and Mikael Vaillant.
\newblock Lethe 1.0: {{An}} open-source parallel high-order computational fluid dynamics software framework for single and multiphase flows.
\newblock \emph{Computer Physics Communications}, 318:\penalty0 109880, January 2026.
\newblock ISSN 0010-4655.
\newblock \doi{10.1016/j.cpc.2025.109880}.

\bibitem[Africa et~al.(2024)Africa, Arndt, Bangerth, Blais, Fehling, Gassm{\"o}ller, Heister, Heltai, Kinnewig, Kronbichler, Maier, Munch, {Schreter-Fleischhacker}, Thiele, Turcksin, Wells, and Yushutin]{africa_2024}
Pasquale~C. Africa, Daniel Arndt, Wolfgang Bangerth, Bruno Blais, Marc Fehling, Rene Gassm{\"o}ller, Timo Heister, Luca Heltai, Sebastian Kinnewig, Martin Kronbichler, Matthias Maier, Peter Munch, Magdalena {Schreter-Fleischhacker}, Jan~P. Thiele, Bruno Turcksin, David Wells, and Vladimir Yushutin.
\newblock The deal.{{II}} library, {{Version}} 9.6.
\newblock \emph{Journal of Numerical Mathematics}, 32\penalty0 (4):\penalty0 369--380, December 2024.
\newblock ISSN 1570-2820, 1569-3953.
\newblock \doi{10.1515/jnma-2024-0137}.

\bibitem[Gassm{\"o}ller et~al.(2018)Gassm{\"o}ller, Lokavarapu, Heien, Puckett, and Bangerth]{gassmoller2018}
Rene Gassm{\"o}ller, Harsha Lokavarapu, Eric Heien, Elbridge~Gerry Puckett, and Wolfgang Bangerth.
\newblock Flexible and {{Scalable Particle-in-Cell Methods With Adaptive Mesh Refinement}} for {{Geodynamic Computations}}.
\newblock \emph{Geochemistry, Geophysics, Geosystems}, 19\penalty0 (9):\penalty0 3596--3604, 2018.
\newblock ISSN 1525-2027.
\newblock \doi{10.1029/2018GC007508}.

\bibitem[Coetzee and Scheffler(2023)]{coetzee_2023a}
Corn{\'e}~J. Coetzee and Otto~C. Scheffler.
\newblock Review: {{The Calibration}} of {{DEM Parameters}} for the {{Bulk Modelling}} of {{Cohesive Materials}}.
\newblock \emph{Processes}, 11\penalty0 (1):\penalty0 5, January 2023.
\newblock ISSN 2227-9717.
\newblock \doi{10.3390/pr11010005}.

\bibitem[Seville et~al.(2000)Seville, Willett, and Knight]{seville_2000}
J.~P.~K. Seville, C.~D. Willett, and P.~C. Knight.
\newblock Interparticle forces in fluidisation: A review.
\newblock \emph{Powder Technology}, 113\penalty0 (3):\penalty0 261--268, December 2000.
\newblock ISSN 0032-5910.
\newblock \doi{10.1016/S0032-5910(00)00309-0}.

\bibitem[Derjaguin et~al.(1975)Derjaguin, Muller, and Toporov]{derjaguin_1975}
B.~V Derjaguin, V.~M Muller, and Yu.~P Toporov.
\newblock Effect of contact deformations on the adhesion of particles.
\newblock \emph{Journal of Colloid and Interface Science}, 53\penalty0 (2):\penalty0 314--326, November 1975.
\newblock ISSN 0021-9797.
\newblock \doi{10.1016/0021-9797(75)90018-1}.

\bibitem[Ai et~al.(2011)Ai, Chen, Rotter, and Ooi]{ai_2011}
Jun Ai, Jian-Fei Chen, J.~Michael Rotter, and Jin~Y. Ooi.
\newblock Assessment of rolling resistance models in discrete element simulations.
\newblock \emph{Powder Technology}, 206\penalty0 (3):\penalty0 269--282, January 2011.
\newblock ISSN 0032-5910.
\newblock \doi{10.1016/j.powtec.2010.09.030}.

\bibitem[Li et~al.(2020)Li, Ucgul, Lee, and Saunders]{li2020a}
Peilin Li, Mustafa Ucgul, Sang-Heon Lee, and Chris Saunders.
\newblock A new approach for the automatic measurement of the angle of repose of granular materials with maximal least square using digital image processing.
\newblock \emph{Computers and Electronics in Agriculture}, 172:\penalty0 105356, May 2020.
\newblock ISSN 01681699.
\newblock \doi{10.1016/j.compag.2020.105356}.

\bibitem[{Beakawi Al-Hashemi} and {Baghabra Al-Amoudi}(2018)]{beakawial-hashemi2018}
Hamzah~M. {Beakawi Al-Hashemi} and Omar~S. {Baghabra Al-Amoudi}.
\newblock A review on the angle of repose of granular materials.
\newblock \emph{Powder Technology}, 330:\penalty0 397--417, May 2018.
\newblock ISSN 0032-5910.
\newblock \doi{10.1016/j.powtec.2018.02.003}.

\bibitem[M{\"u}ller et~al.(2021)M{\"u}ller, Fimbinger, and Brand]{muller2021a}
Dominik M{\"u}ller, Eric Fimbinger, and Clemens Brand.
\newblock Algorithm for the determination of the angle of repose in bulk material analysis.
\newblock \emph{Powder Technology}, 383:\penalty0 598--605, May 2021.
\newblock ISSN 00325910.
\newblock \doi{10.1016/j.powtec.2021.01.010}.

\bibitem[Katagiri et~al.(2026)Katagiri, Shoji, Yoneda, and Takeya]{katagiri2026}
Jun Katagiri, Nana Shoji, Jun Yoneda, and Satoshi Takeya.
\newblock Limits of identifiability and uncertainty propagation in {{DEM}} parameter calibration using the angle of repose as the objective function.
\newblock \emph{Computational Particle Mechanics}, 14:\penalty0 334--345, April 2026.
\newblock ISSN 2196-4386.
\newblock \doi{10.1016/j.cpms.2026.03.008}.

\end{thebibliography}
